\documentclass[12pt]{article}
\usepackage{graphicx,epsfig}
\usepackage{amsmath,amssymb}
\usepackage{array}
\usepackage{arydshln}
\usepackage{color}
\usepackage{cite}

\newcommand{\be}{\begin{equation}}
\newcommand{\ee}{\end{equation}}
\newcommand{\bea}{\begin{eqnarray}}
\newcommand{\eea}{\end{eqnarray}}
\newcommand{\beq}{\begin{equation}}
\newcommand{\eeq}{\end{equation}}
\newcommand{\beqa}{\begin{eqnarray}}
\newcommand{\eeqa}{\end{eqnarray}}
\newcommand{\ba}{\begin{array}}
\newcommand{\ea}{\end{array}}

\newcommand{\cL}{{\cal L}}
\newcommand{\cH}{{\cal H}}

\newcommand{\cO}{{\cal O}}
\newcommand{\cM}{{\cal M}}
\newcommand{\ord}[1]{\mathcal{O}({#1})}

\newcommand{\lsim}{\stackrel{<}{_\sim}}
\newcommand{\no}{\nonumber}
\definecolor{nicered}{rgb}{0.7,0.1,0.1}
\title{\Large\bf\boldmath
$U(2)$ and Minimal Flavour Violation in Supersymmetry
}
\date{}
\author{Riccardo Barbieri$^{a}$, Gino Isidori$^{b}$, 
Joel Jones-P\'erez$^{b}$, \\ Paolo Lodone$^{a}$, 
 David M. Straub$^{a}$
\\[0.5 cm]
$^a${\em\normalsize Scuola Normale Superiore and INFN, Piazza dei Cavalieri 7, 56126 Pisa, Italy} \\
$^b${\em\normalsize INFN, Laboratori Nazionali di Frascati, Via E.~Fermi 40, I-00044 Frascati, Italy}
}

\begin{document}

\maketitle

\begin{abstract}
\noindent
Rather than sticking
to the full $U(3)^3$ approximate symmetry normally invoked in Minimal Flavour Violation, we analyze the consequences on the current flavour data of a suitably broken $U(2)^3$ symmetry acting on the first two generations of quarks and squarks. A definite correlation emerges between the $\Delta F=2$ amplitudes $\cM( K^0 \to \bar K^0 )$, $\cM( B_d \to \bar B_d )$ and $\cM( B_s \to \bar B_s )$, which can resolve the current tension between  between $\cM( K^0 \to \bar K^0 )$ and $\cM( B_d \to \bar B_d )$, while  predicting $\cM( B_s \to \bar B_s )$. In particular, the CP violating asymmetry in  $B_s \to \psi \phi$
is predicted to be positive $S_{\psi\phi} = 0.12 \pm 0.05$
and above its Standard Model value ($S_{\psi\phi} = 0.041\pm0.002$).
The preferred region for the gluino and the left-handed sbottom masses is below 
about $1\div 1.5$~TeV. An existence proof of a dynamical model realizing the $U(2)^3$ 
picture is outlined.
\end{abstract}

\section{Introduction}
Explaining  the masses and mixings of quarks and leptons remains a fundamental open problem in particle physics. What the last decade of experimental developments has added to this problem is the evidence that the CKM picture of the quark flavours, as realized in the Standard Model, is fundamentally at work. 

How does weak-scale supersymmetry confront these statements? While such a question  is relevant for any extension of the SM, it is especially pregnant in the case of supersymmetry, which doubles the number of  flavoured degrees of freedom    at the Fermi scale with their own masses and mixings. In a sense this appears to be both a special problem and an opportunity. It is a special problem because, with squarks in the hundreds of GeVs, the preservation 
of the CKM picture to a sufficient level of accuracy is non trivial. It is an opportunity because the deviations from a strict CKM picture that should show up at some level might bring new key information to attack the problem of the origin of flavour breaking at all. Also in view of the  tension that emerges from the cumulative fits of flavour physics in the strict SM, this motivates us to reconsider the flavour problem in supersymmetry.

A phenomenological ``near-CKM'' picture of flavour physics is highly suggestive of a suitable flavour symmetry approximately operative on the entire supersymmetric extension of the SM, whatever it may be. 
 Among the symmetries that have been considered, two are of interest here:
\begin{itemize}
\item $U(3)_Q\times U(3)_u\times U(3)_d$, broken by {\it spurions} transforming as $Y_u= (3, \bar{3}, 1)$ and $Y_d= (3, 1, \bar{3})$ \cite{Chivukula:1987py,Hall:1990ac,D'Ambrosio:2002ex};
\item $U(2)$ acting on the first two generations of quark superfields (and commuting with the gauge group), broken by one  single doublet and by one or more rank-two 
tensors~\cite{Pomarol:1995xc,Barbieri:1995uv}.
\end{itemize}
The first case -- $U(3)^3$ for brevity -- corresponds to the standard 
Minimal Flavour Violation (MFV) hypothesis and can result from gauge mediation of supersymmetry breaking.  $U(3)^3$ can explain the lack of flavour signals so far from s-partner exchanges, provided one take small enough  flavour-blind CP-phases to cope with the limits from the Electric Dipole Moments (the so called supersymmetric CP-problem). This in turn hampers the possible interpretation of the recently measured CP-asymmetries in the $B$-system 
in terms of new physics. No attempt is made to address the ``fermion mass problem''.

A step in this direction is instead taken in the second case, based on the strong hierarchical pattern of the Yukawa couplings with only one of them, or two at most, of order unity. In Ref.s~\cite{Pomarol:1995xc,Barbieri:1995uv}  this pattern is assumed to result from a weakly broken $U(2)$ symmetry acting on the first two generations of quarks superfields consistently with $SU(3)\times SU(2)\times U(1)$ gauge invariance. $U(2)$ 
can also go  a long way in explaining the absence, so far, of new flavour changing phenomena, with the special feature, not allowed in the $U(3)^3$ case, that the first two generations of squarks can be significantly heavier than the third generation ones.
This is crucial to solve the supersymmetric CP problem, making compatible sizable flavour-blind CP phases with the current limits on the Electric Dipole Moments. However, a $U(2)$ symmetry acting
on both left- and right-handed fields, does not provide in general 
a sufficient protection of flavour-violating 
effects in the right-handed sector, which are strongly constrained by present data.

\section{Definition of the framework}

For reasons that will be clear shortly, here we consider  an approximate $U(2)_Q\times U(2)_u\times U(2)_d$ flavour symmetry, intermediate between the two previous cases and still motivated by the pattern of quark masses and mixings. Furthermore,  in analogy with the MFV case, we assume that this $U(2)^3$ is broken by {\it spurions} transforming as $\Delta Y_u= (2, \bar{2}, 1)$ and 
$\Delta Y_d= (2, 1, \bar{2})$. In fact, if these {\it bi-doublets} were the only breaking parameters, the third generation, made of singlets under $U(2)^3$, would not be able to communicate with the first two generations at all. For this to happen, one needs single doublets, at least one,  under any  of the three $U(2)$'s. The only such doublet that can explain the natural size of the  quark masses and mixings, up to factors of order unity, transforms under $U(2)_Q\times U(2)_u\times U(2)_d$ as $V = (2,1,1)$.

Combining the various symmetry breaking terms, as described in Appendix A,
the standard $3\times 3$ Yukawa matrices in generation space end up with the following form:
\begin{align}
Y_u= y_t \left(\begin{array}{c:c}
 \Delta Y_u & x_t\,V \\\hdashline
 0 & 1
\end{array}\right), & &
Y_d= y_b \left(\begin{array}{c:c}
 \Delta Y_d & x_b\,V \\\hdashline
 0 & 1
\end{array}\right),
\label{yukawa}
\end{align}
where $\Delta Y_u$ and $\Delta Y_d$ have been suitably rescaled, $y_t, y_b$ are the third generation Yukawa couplings and $x_t, x_b$ are complex parameters of $\ord{1}$. The $2\times 2$ matrices $\Delta Y_u$ and $\Delta Y_d$ and the vector $V$ are the small symmetry breaking parameters of $U(2)_Q\times U(2)_u\times U(2)_d$ with entries of order $\lambda^2$ or smaller, with $\lambda$ the sine of the Cabibbo angle.
Analogous expressions, detailed in Appendix B, hold for the three soft mass matrices $m^2_{\tilde{Q}}, m^2_{\tilde{u}}, m^2_{\tilde{d}}$. 

By suitable unitary transformations one can go to the physical basis for quarks and squarks with the consequent appearance of mixing matrices in the various interaction terms, in particular the standard charged current interactions and the gaugino interactions of the down quark-squarks
\begin{equation}
(\bar{u}_L\gamma_\mu V_{CKM} d_L) W_\mu~,~~
(\bar{d}_{L,R} W^d_{L,R} \tilde{d}_{L,R}) \tilde{g}~.
\end{equation}
As shown in Appendixes A and B, to a good approximation the matrices $V_{CKM}$ and $W^d_L$ have the following correlated forms
\be
 V_{\rm CKM}=\left(\begin{array}{ccc}
 1- \lambda^2/2 &  \lambda & s_u s e^{-i \delta}  \\
-\lambda & 1- \lambda^2/2   & c_u s  \\
-s_d s \,e^{i (\phi+\delta)} & -s c_d & 1 \\
\end{array}\right),
\label{CKM}
\ee
\bea
W^d_L = \left(\begin{array}{ccc}
 c_d &  s_d  e^{-i(\delta +\phi)}  & -s_d s_L e^{i\gamma} e^{-i(\delta +\phi)}  \\
-s_d e^{i(\delta +\phi)}  &  c_d & -c_d s_L e^{i\gamma}   \\
  0  &  s_L e^{-i\gamma} & 1 \\
\end{array}\right),~
\label{WL}
\eea
where the phases $\phi$ and $\delta$ are related to each other and to 
the real and positive parameter $\lambda$ via
\be
s_uc_d - c_u s_d e^{-i\phi}  = \lambda e^{i \delta}~,
\ee
the real parameter $s_L$ is of order $\lambda^2$, and $\gamma$ is an independent 
CP-violating phase. At the same time the off-diagonal entries of 
the matrix $W^d_R$ are negligibly small.

Up to phase redefinitions, eq.s (\ref{CKM}) and (\ref{WL}) were obtained in 
Ref.~\cite{Barbieri:1997tu} based on a $U(2)$ symmetry. There, however, with a single $U(2)$ not distinguishing between left and right, a mixing matrix $W^d_R$ was also present involving a new mixing angle 
($s_L \rightarrow s_R$) and a new phase ($\gamma \rightarrow \gamma_R$). 
As it will be apparent in the next Section, the simultaneous 
presence of $W^d_L$ and $W^d_R$  would lead to a $\Delta S=2$ $L R$ operator, 
which  corrects by a too large amount the CP-violating $\epsilon_K$ parameter 
due to its chirally enhanced matrix element.

\section{Implications}

While the CKM picture has been very successful in describing experimental 
observations of flavour and CP violation, recently there are mounting tensions 
in this description. Firstly, there is a tension among the CP violating 
parameter in the $K$ system $|\epsilon_K|$, the ratio of mass differences in 
the $B_{d,s}$ systems $\Delta M_d/\Delta M_s$ and the mixing induced CP 
asymmetry in $B_d\to\psi K_S$, which in the SM measures $\sin(2\beta)$ 
\cite{Lunghi:2008aa,Buras:2008nn,Altmannshofer:2009ne,Lunghi:2010gv,Bevan:2010gi}. Secondly, there 
are hints for a sizable mixing-induced CP asymmetry in $B_s\to J/\psi\phi$, 
implying non-standard CP violation in the $B_s$ system, and an anomalous 
dimuon charge asymmetry, pointing to non-standard CP violation either in the 
$B_d$ or $B_s$ systems \cite{Abazov:2010hv,Lenz:2010gu}.

To expose the tension among $|\epsilon_K|$,  $\Delta M_d/\Delta M_s$ and 
$\sin(2\beta)$ in the SM, we perform a global fit of the CKM matrix. By removing one observable from the fit, one obtains a prediction for it which can then be compared to its experimental value.

In a second step, we discuss the modification of the $\Delta F=2$ observables entering the fit in the $U(2)^3$ setup. Performing a fit of the CKM matrix and the relevant additional model parameters allows us to predict the size of CP violation in the $B_s$ system and to get information on the scale of sparticle masses. 

\subsection{Input data and Standard Model fit}
\label{sec:SMfit}

\begin{table*}
\renewcommand{\arraystretch}{1.3}
 \begin{center}
\begin{tabular}{|lll|lll|}
\hline
$|V_{ud}|$ & $0.97425(22)$ &\cite{Hardy:2008gy}& $f_K$  & $(155.8\pm1.7)$ MeV & \cite{Laiho:2009eu}\\
$|V_{us}|$ & $0.2254(13)$ &\cite{Antonelli:2010yf}& $\hat B_K$ & $0.724\pm0.030$ &\cite{Colangelo:2010et} \\
$|V_{cb}|$ & $(40.89\pm0.70)\times10^{-3}$ &\cite{Lenz:2010gu}& $\kappa_\epsilon$ & $0.94\pm0.02$ & \cite{Buras:2010pza}\\
$|V_{ub}|$ & $(3.97\pm0.45)\times10^{-3}$ &\cite{Beauty11}& $f_{B_s}\sqrt{\hat B_s}$  & $(291\pm16)$ MeV &\cite{Lunghi:2011xy}\\
$\gamma_{\rm CKM}$ & $(74\pm11)^\circ$ &\cite{Bevan:2010gi}& $\xi$ & $1.23\pm0.04$ &\cite{Lunghi:2011xy}\\
$|\epsilon_K|$ & $(2.229\pm0.010)\times10^{-3}$ &\cite{Nakamura:2010zzi} &&&\\ 
$S_{\psi K_S}$ & $0.673\pm0.023$ &\cite{Asner:2010qj} &&&\\
$\Delta M_d$ & $(0.507\pm0.004)\,\text{ps}^{-1}$ &\cite{Asner:2010qj} &&&\\
$\Delta M_s$ & $(17.77\pm0.12)\,\text{ps}^{-1}$ &\cite{Abulencia:2006ze} &&&\\
\hline
 \end{tabular}
 \end{center}
\caption{Observables and hadronic parameters used as input to the CKM fit.}
\label{tab:inputs}
\end{table*}

We perform global fits of the Wolfenstein CKM parameters $\lambda$, $A$, $\bar\rho$ and $\bar\eta$ to (a subset of) the observables given in the left column of table~\ref{tab:inputs}. To this end, we use a Markov Chain Monte Carlo with the Metropolis algorithm to determine the Bayesian posterior probability distribution for the input parameters. We treat all errors as Gaussian, taking into account the experimental uncertainties indicated in the left column of table~\ref{tab:inputs} as well as the theoretical ones, due to the hadronic parameters collected in the right column of table~\ref{tab:inputs}.

\begin{figure}[tbp]
\begin{center}
\includegraphics[width=0.9\textwidth]{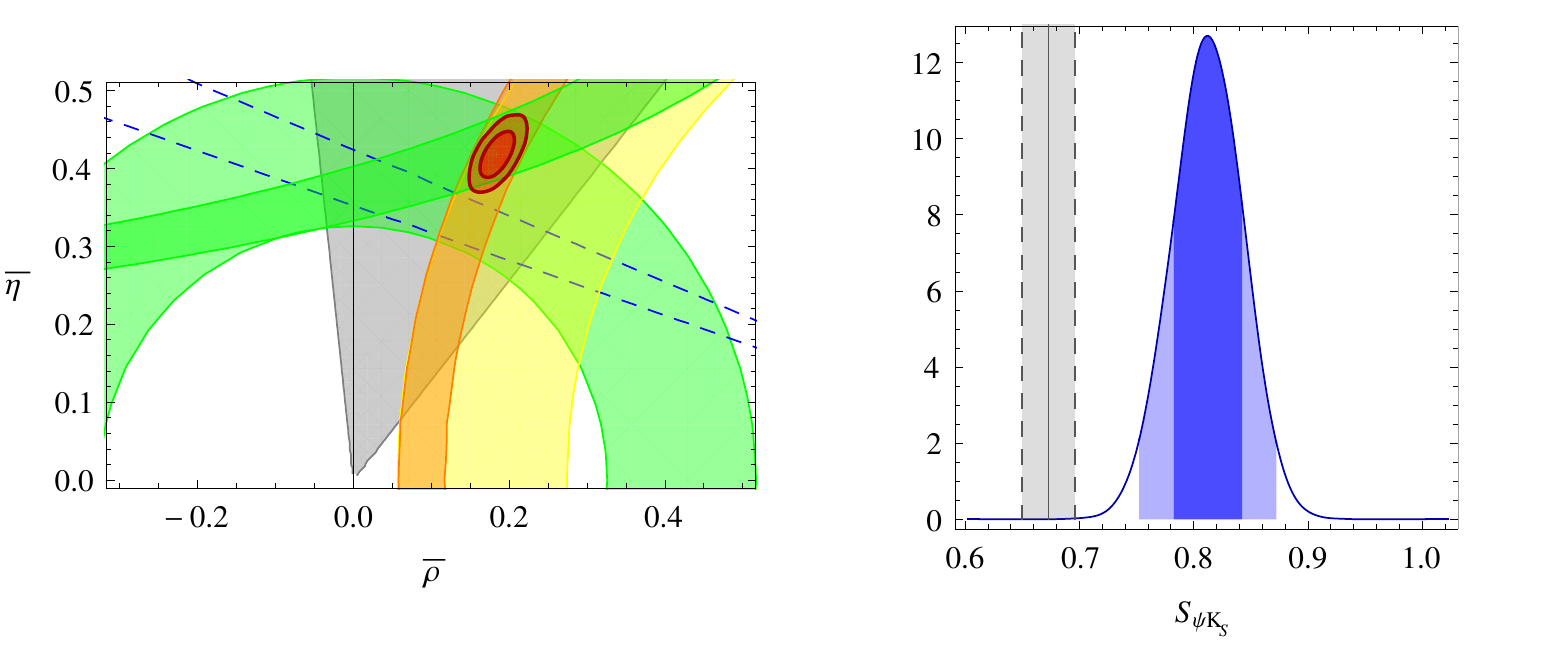}
\includegraphics[width=0.9\textwidth]{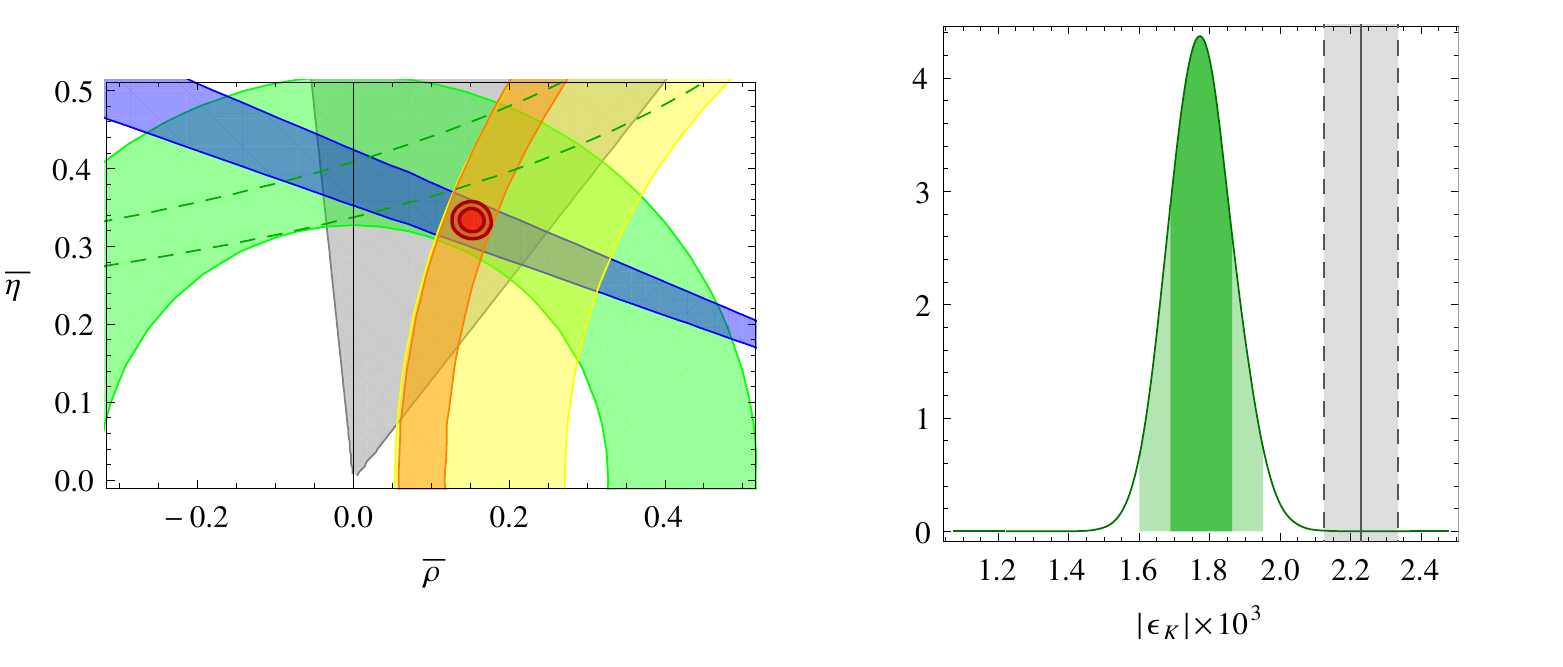}
\end{center}
\caption{\normalsize Results of two global fits of the CKM matrix using tree-level and $\Delta F=2$ observables, excluding $S_{\psi K_S}=\sin(2\beta)$ (top row) or $|\epsilon_K|$ (bottom row). The bands in the left panels correspond to $2\sigma$ errors. The dotted bands in the right panels correspond 
to $1\sigma$ errors. }
\label{fig:smfit}
\end{figure}

Figure~\ref{fig:smfit} shows the fit results for two fits: in the first case all the constraints in the left column of table~\ref{tab:inputs} {\em except for $S_{\psi K_S}=\sin(2\beta)$} have been used, in the second case all constraints  {\em except for $|\epsilon_K|$}. The left panels show the $2\sigma$ bands of the individual constraints in the $(\bar\rho,\bar\eta)$ plane and the 68\% and 95\% C.L. region for $\bar\rho$ and $\bar\eta$. The dashed lines show the band of the ``unused'' constraint, which in both cases clearly deviates from the region preferred by the fit. This becomes even more apparent comparing the probability density functions of $S_{\psi K_S}$ and  $|\epsilon_K|$ to their experimental values as shown in the right panels.

These tensions, if explained by NP, could be due to non-standard contributions in neutral kaon mixing, non-standard CP violation in $B_d$ mixing or a non-universal modification of the mass differences in the $B_d$ and $B_s$ systems. In the next section, we will discuss which of these solutions is possible in the $U(2)^3$ setup.

\subsection{Supersymmetric fit}
\label{sec:SUSYfit}

\begin{figure}[tbp]
\begin{center}
\includegraphics[width=0.48\textwidth]{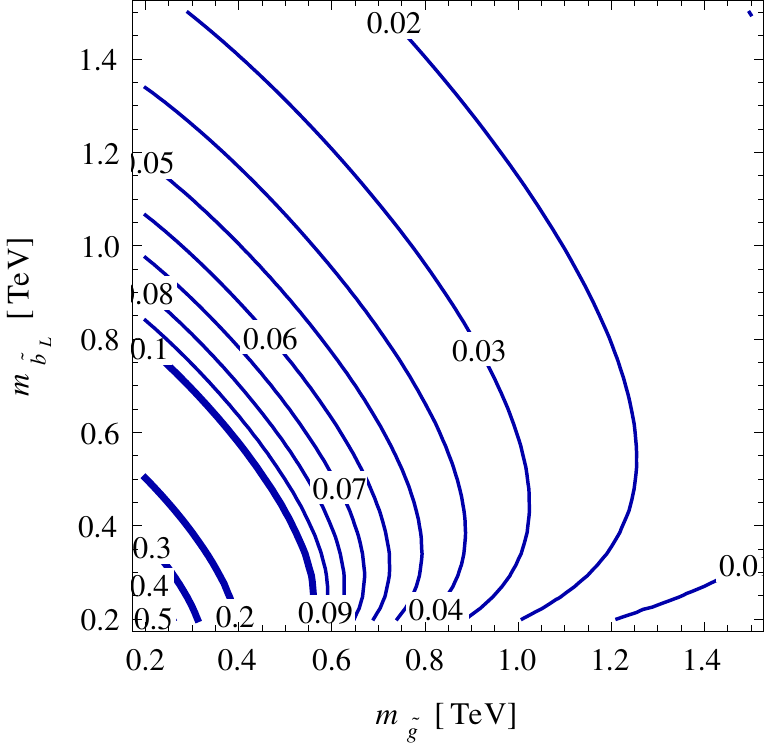}
\end{center}
\caption{Value of the loop function $F_0$ defined in (\ref{eq:F0}) as function of the gluino and left-handed sbottom masses.}
\label{fig:F0}
\end{figure}

For $m_{Q_l}^2 \ll m_{Q_h}^2$
the three down-type $\Delta F=2$ amplitudes,
including SM and gluino-mediated 
contributions, assume the  following simple form
\bea
\cM( K^0 \to \bar K^0 ) &=& \left|\cM^{(\rm tt)}_{\rm SM}\right| 
\frac{(V_{ts}^*V_{td})^2}{|V_{ts}^*V_{td}|^2} 
\left[1+ \frac{s_L^4 c^4_d}{|V_{ts}|^4 }\, F_0~ \right] 
+ \cM_{\rm SM}^{(\rm tc+cc)}~, \label{eq:MKK}\\ 
\cM( B_d \to \bar B_d ) &=& \left|\cM_{\rm SM}\right| 
\frac{(V_{tb}^*V_{td})^2}{|V_{tb}^*V_{td}|^2} 
\left[1+ \frac{s_L^2 c^2_d}{|V_{ts}|^2 } e^{-2i\gamma}\, F_0~ \right]~, 
\label{eq:BdBd}\\ 
\cM( B_s \to \bar B_s ) &=& \left|\cM_{\rm SM}\right| 
\frac{(V_{tb}^*V_{ts})^2}{|V_{tb}^*V_{ts}|^2} 
\left[1+ \frac{s_L^2 c^2_d}{|V_{ts}|^2 } e^{-2i\gamma}\, F_0~ \right]~, 
\label{eq:MBsBs}
\eea
where in the kaon case we have separated the leading top-top contribution from the 
subleading top-charm and charm-charm terms.
The function $F$ is 
\bea 
F_0 &=& \frac{2}{3} \left(\frac{g_s}{g} \right)^4 \frac{ m_W^2 }{ m^2_{Q_3} } \frac{1}{S_0(x_t)}
\left[ f_{0}(x_g) + \cO\left( \frac{m_{Q_l}^2}{m_{Q_h}^2} \right)\right]~, 
\label{eq:F0} \\
f_{0}(x) & = & \frac{11 + 8 x -19x^2 +26 x\log(x)+4x^2\log(x)}{3(1-x)^3}~,  \\
x_g &=& \frac{ m^2_{\tilde g} }{ m_{Q_3}^2 }~,
\qquad\quad\  f_{0}(1)=1~,  \no
\eea
where $S_0(x_t = m_t^2/m_W^2)\approx 2.4$ is the SM one-loop 
electroweak coefficient function. Note that, since SM and glu\-ino-mediated 
contributions generate the same $\Delta F=2$ effective operator,
all non-perturbative effects and long-distance QCD corrections 
have been factorized.
The typical size of $F_0$, as a function of the gluino mass and $m_{Q_3}$, is shown in figure~\ref{fig:F0}.

Eq.s~(\ref{eq:MKK})--(\ref{eq:MBsBs}) lead to remarkably simple expressions for the modification of the $\Delta F=2$ observables entering the CKM fit. Defining
$x={s_L^2 c^2_d/|V_{ts}|^2}$,
one can write
\begin{align}
\epsilon_K&=\epsilon_K^\text{SM(tt)}\times\left(1+x^2F_0\right) +\epsilon_K^\text{SM(tc+cc)} ~
\label{eq:epsKxF}\\
S_{\psi K_S} &=\sin\left(2\beta + \text{arg}\left(1+xF_0 e^{2i\gamma}\right)\right) ~,\label{eq:Spk} \\
\Delta M_d &=\Delta M_d^\text{SM}\times\left|1+xF_0 e^{2i\gamma}\right| ~,
\label{eq:DMdxF}\\
\frac{\Delta M_d}{\Delta M_s} &= \frac{\Delta M_d^\text{SM}}{\Delta M_s^\text{SM}} ~.
\label{eq:MdMs}
\end{align}
Analogously, the mixing-induced CP asymmetry in $B_s\to J/\psi\phi$ can be written as
\begin{equation}
S_{\psi\phi} =\sin\left(2|\beta_s| - \text{arg}\left(1+xF_0 e^{2i\gamma}\right)\right) ~,
\label{eq:SpsiphixF}
\end{equation}
where $\beta_s=-\text{Arg}\left[-(V_{ts}^*V_{tb})/(V_{cs}^*V_{cb})\right]$ is the SM mixing phase.

We are now in a position to perform a fit of the CKM matrix in the supersymmetric case, varying $x$, $F_0$ and $\gamma$ of eq.s (\ref{eq:epsKxF})--(\ref{eq:DMdxF}) in addition to the 4 Wolfenstein parameters and using all observables in the left column of table~\ref{tab:inputs} as constraints, with the appropriate modification of $\Delta F=2$ amplitudes as discussed above. Given the typical size of $F_0$ as shown in figure~\ref{fig:F0}, we impose a flat prior $F_0<0.2$.

\begin{figure}[tbp]
\begin{center}
\includegraphics[width=0.91\textwidth]{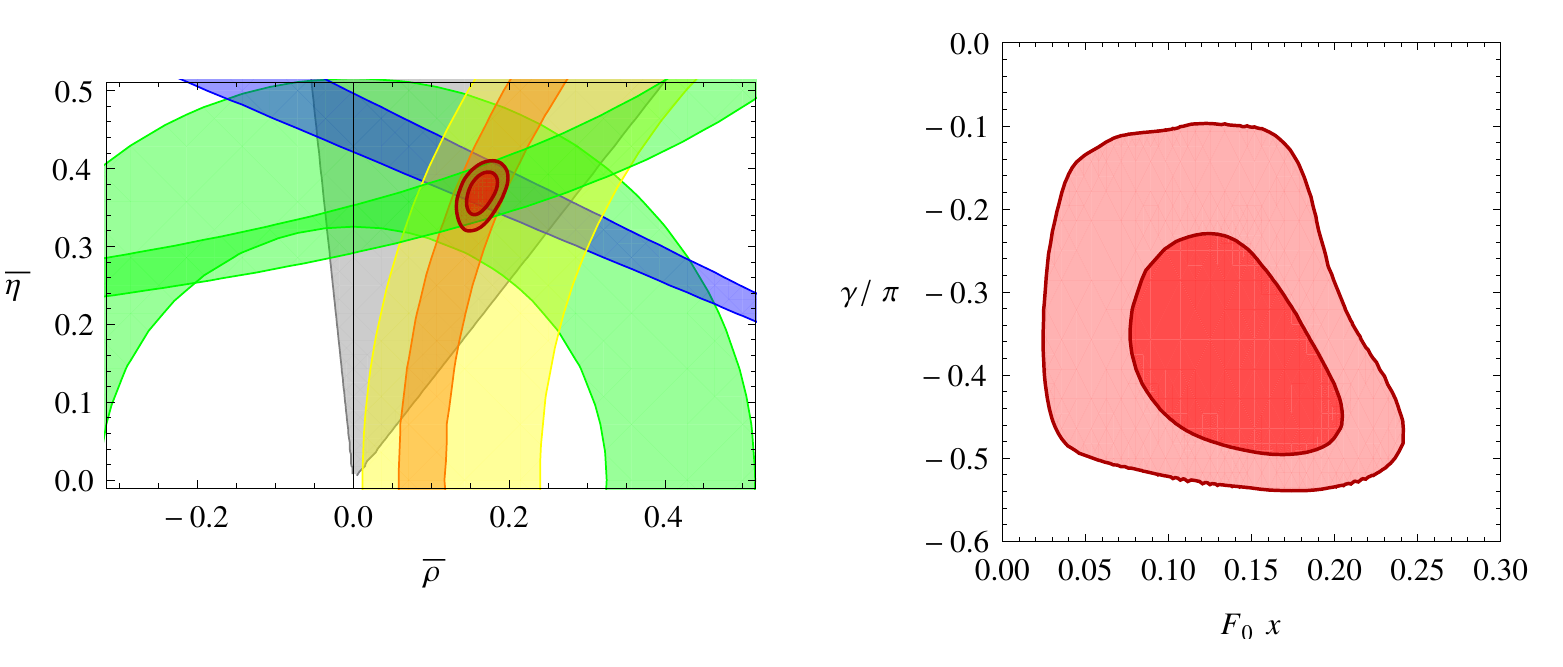}
\end{center}
\caption{{\em Left: result of the global fit with inclusion of the corrections as in eq.s (\ref{eq:epsKxF}-\ref{eq:DMdxF}). Right: preferred values of the parameters defined in text, as determined from the fit.}}
\label{fig:u2fit}
\end{figure}

\begin{figure}[tbp]
\begin{center}
\includegraphics[width=0.9\textwidth]{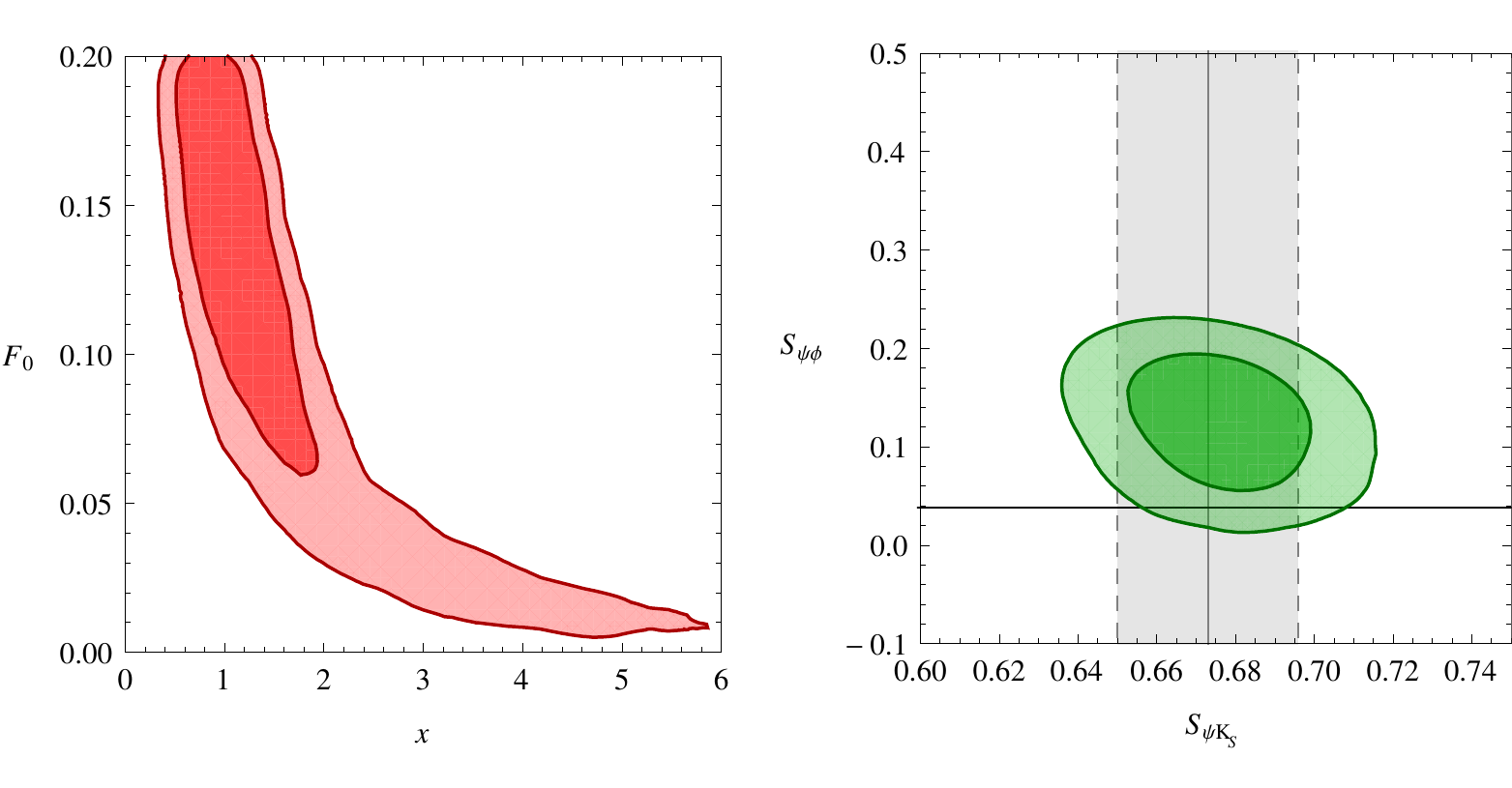}
\end{center}
\caption{Correlation among the preferred values of $x$ and $F_0$ (left) 
and prediction of   $S_{\psi\phi}$ as a 
function of $S_{\psi K_S}$ (right) as determined from the supersymmetric fit.}
\label{fig:spsiphi}
\end{figure}

The left panel of figure~\ref{fig:u2fit} shows the fit result in the $(\bar\rho,\bar\eta)$ plane. The tension among $|\epsilon_K|$, $S_{\psi K_S}$ and $\Delta M_d/\Delta M_s$ has disappeared. More precisely, 
we find $\chi^2/{\rm N_{dof}}=0.7/2$, compared to $\chi^2/{\rm N_{dof}}=9.8/5$ for the full SM fit.
This is due both to a positive SUSY contribution to $|\epsilon_K|$ as well as a new phase 
in $B_d$ mixing. Note that the positive sign of the SUSY contribution to $|\epsilon_K|$
is an unambiguous prediction of our framework.
The right panel shows the preferred values for the combination $F_0 x$ and the phase $\gamma$ entering $B_{d,s}$ mixing.
As shown in the left panel of figure~\ref{fig:spsiphi}, $F_0$ and $x$ are not very well constrained separately, but $F_0\gtrsim0.05$ is preferred by the fit, implying sub-TeV gluino and squark masses 
(see figure~\ref{fig:F0}).

The non-zero value of $\gamma$ required by the fit to solve the CKM tensions implies non-standard CP-violation in the $B_s$ system by means of equation~(\ref{eq:SpsiphixF}). We show the fit prediction for $S_{\psi\phi}$ in the right panel of figure~\ref{fig:spsiphi} in the $S_{\psi K_S}$ vs.~$S_{\psi\phi}$ plane. While  $S_{\psi K_S}$ coincides with the experimental measurement (note that it was among the fit constraints),  $S_{\psi\phi}$ is clearly preferred to be larger than its tiny SM value, indicated by a horizontal line. 
The pattern implied by (\ref{eq:epsKxF}-\ref{eq:MdMs}) was already noticed in \cite{Barbieri:1998qs} assuming the dominance of the LL operators.
The correlation between  $S_{\psi K_S}$ and $S_{\psi\phi}$ implied by eq.s~(\ref{eq:Spk}) 
and (\ref{eq:SpsiphixF}) is the same pointed out in~\cite{Ligeti:2010ia}
in the context of effective theory approaches 
with a horizontal $SU(2)$ symmetry acting on left-handed light quarks.

For the semi-leptonic asymmetry $a^s_\text{SL}$ and the like-sign dimuon 
charge asymmetry $A_\text{SL}$ measured at the Tevatron, we find values below 
the permille level. We note that an enhancement of $A_\text{SL}$ above the 3 
permille level requires NP contributions to the absorptive part of the mixing 
amplitude \cite{Blanke:2006ig}, which are not generated in our framework.

\section{\boldmath  A dynamical model}

As recalled in Section 1, the $U(3)^3$ of MFV arises in gauge mediation of supersymmetry breaking. The only condition for this to be the case is that the scale at which the Yukawa couplings are generated is higher than the scale of the messenger masses. Here we briefly outline a possible dynamical model for the generation of the $U(2)^3$ described above, suitably modifying a proposal in Ref. \cite{Craig:2011yk}.

The basic splitting between the third and the first two generations arises from a doubling of the SM gauge group, $G_1^\text{SM}$ and $G_2^\text{SM}$, a ``two-site deconstruction'' of the SM, with the third generation of matter superfields transforming in the usual way under $G_1^\text{SM}$ and the first two generations  under $G_2^\text{SM}$. Like the third generation of matter, the two Higgs doublets of the MSSM also transform only under $G_1^\text{SM}$. As usual, the SM is made to emerge at low energy by spontaneously breaking the product group $G_1^\text{SM} \times G_2^\text{SM}$ down to the diagonal subgroup by the VEV's of vector-like link fields. Departing from the assignments in \cite{Craig:2011yk}, we choose their transformation properties  
as indicated in table~\ref{tab:link},
with an eye to the desired pattern of the Yukawa couplings in flavour space. To avoid unwanted light states the superpotential will have to include quartic terms of the 
form $\chi_h \chi_\ell \bar{\chi}_h \bar{\chi}_\ell$.
The definition of the model is complete by including a supersymmetry breaking sector directly coupled to messengers of mass $M$ charged under $G_2^\text{SM}$.

\begin{table}[t]
\begin{center}
\renewcommand{\arraystretch}{1.2}
\begin{tabular}{|c||c|c|}
\hline
Chiral field & $G^\text{SM}_{1}$ & $G^\text{SM}_{2}$ \\ \hline
$\chi_h$ & $(3,2,\frac{1}{6})$ & $(\overline{3},2,-\frac{1}{6})$ \\ \hline
$\tilde{\chi}_h$ & $(\overline{3},2,-\frac{1}{6})$ & $({3},2,\frac{1}{6})$ \\ \hline
$\chi_\ell$ & $(1,2,\frac{1}{2})$ & $(1,2,-\frac{1}{2})$ \\ \hline
$\tilde{\chi}_\ell$ & $(1,2,-\frac{1}{2})$ & $(1,2,\frac{1}{2})$ \\ \hline
\end{tabular} 
\renewcommand{\arraystretch}{1.0}
\caption{Transformation properties of the link fields under $G_1^\text{SM} \times G_2^\text{SM}$.
\label{tab:link}}
\end{center}
\end{table}

In absence of Yukawa couplings such a model has a built in $U(2)^3$ flavour symmetry with the first and second generation sfermions receiving a standard gauge-mediation two loop mass
\begin{equation}
m^2_{GM} \approx \left(\frac{\alpha}{4\pi}\right)^2 \left(\frac{F}{M}\right)^2,
\end{equation}
while the third generation sfermion masses are suppressed 
since they effectively come from gaugino mediation
\begin{equation}
m^2_{gM} \approx \left(\frac{\alpha}{4\pi}\right)^3  \left(\frac{F}{M}\right)^2.
\end{equation}
Let us now introduce the most general Yukawa interactions of lowest possible dimensionality, weighted by inverse powers of a mass scale $M^*$, not necessarily related to $M$, but otherwise with dimensionless couplings of order unity. Defining the small parameters
\begin{equation}
\epsilon_\ell \equiv \frac{\langle\chi_\ell\rangle}{M^*} = \frac{\langle\tilde{\chi}_\ell\rangle}{M^*},~~~
\epsilon_h \equiv \frac{\langle\chi_h\rangle}{M^*} = \frac{\langle\tilde{\chi}_h\rangle}{M^*},
\end{equation}
one finds the following textures for the quark and squark mass matrices:
\begin{equation}
Y_u ,\, Y_d \sim
\left(
\begin{array}{ccc}
\epsilon_\ell & \epsilon_\ell & \epsilon_h \\
\epsilon_\ell & \epsilon_\ell & \epsilon_h \\
\epsilon_\ell\epsilon_h & \epsilon_\ell\epsilon_h & 1 \\
\end{array}
\right)~, 
\end{equation}
\begin{equation}
m^2_{\tilde{u}} \sim m^2_{\tilde{d}} \sim
\left(
\begin{array}{ccc}
m^2_{GM} & 0 & \epsilon_\ell \epsilon_h m^2_{GM} \\
0 & m^2_{GM} & \epsilon_\ell \epsilon_h m^2_{GM} \\
\epsilon_\ell\epsilon_h m^2_{GM} & \epsilon_\ell\epsilon_hm^2_{GM} & m^2_{\tilde{g}M} \\
\end{array}
\right)~,
\ee
\be
m^2_{\tilde{Q}} \sim
\left(
\begin{array}{ccc}
m^2_{GM} & 0 &  \epsilon_h m^2_{GM} \\
0 & m^2_{GM} &  \epsilon_h m^2_{GM} \\
\epsilon_h m^2_{GM} & \epsilon_hm^2_{GM} & m^2_{\tilde{g}M} \\
\end{array}
\right) \, ,
\end{equation}
where we have neglected $\epsilon^2_h$ terms in 
the $31$ and $32$ entries of $Y_d$ and in $m^2_{\tilde{d}}$.

Taking $\epsilon_\ell \approx \epsilon_h \approx 10^{-2}$, these matrices are approximately equivalent to the ones described in Section 2. Neglecting the small $\epsilon_\ell \epsilon_h$ terms, the main difference is the absence of correlation between the entries $i3, i=1,2$ of the matrices $Y_u, Y_d, m^2_{\tilde{Q}}$ implied in the general analysis by assuming the presence of a single spurion doublet $V$ under $U(2)_Q$. As an example, such correlation can be effectively obtained here by forcing  vanishing $13$ entries in these matrices via a discrete symmetry
\begin{eqnarray*}
& (Q, \, \overline{u},\, \overline{d})_2 ,\, \chi_h,\, \tilde{\chi}_h,\, \chi_{\ell 1},\, \tilde{\chi}_{\ell 1} \rightarrow 
 -(Q, \, \overline{u},\, \overline{d})_2 ,\, -\chi_h,\, -\tilde{\chi}_h,\,- \chi_{\ell 1},\, -\tilde{\chi}_{\ell 1}
\end{eqnarray*}
while all the other fields are untouched.
The only extra ingredient with respect to the minimal model is an additional link field $\chi_{\ell 2}\, , \tilde{\chi}_{\ell 2}$ with the same quantum numbers of $\chi_{\ell 1},\, \tilde{\chi}_{\ell 1}$ but neutral under $Z_2$. Other discrete symmetries, which may be worth studying,
can be invoked to justify mass splitting and mixing angles of the
first two generations.

\section{\boldmath General consequences of $U(2)^3$ and MFV}

Within the supersymmetric framework we are considering,
gluino-mediated amplitudes are the dominant 
non-standard effect in $\Delta F=2$ observables. However, it is worth to 
stress that the results of the fit in 
Section.~\ref{sec:SUSYfit} are, to a large extent, valid
also beyond the assumption of gluino-mediated dominance and even 
beyond supersymmetry: they are a general consequences of $U(2)^3$
and its breaking pattern.

Employing a general effective-theory approach, 
we can analyse the general structure of FCNC amplitudes 
by considering  higher-dimensional operators
formally invariant under $U(2)_{Q}\times U(2)_{u}\times U(2)_{d}$, 
constructed from SM fields and $U(2)^3$-breaking spurions. 
As in the MFV case~\cite{D'Ambrosio:2002ex}, in our framework 
the leading flavour-changing amplitudes 
are of left-handed type and, to a good approximation, 
can be evaluated neglecting the effects of 
light-quark masses (i.e. setting $\Delta Y_{u,d} \to 0$).

The generic combination of left-handed quark bilinears up to $\cO(\epsilon^2)=\cO(\lambda^4)$ 
has the following structure, 
\bea
\hat X_{LL} = \bar q_i X_{ij} q_j =
a_1  \bar Q^\dagger Q + a_3 \bar q_3 q_3
+ b_{13} (\bar Q V^*)  q_3 + b_{31} \bar q_3 (V^T Q) 
+ a_2 (\bar Q V^*) (V^T Q)  +\cO(\epsilon^3)~,
\eea
where $a_i$ and $b_{ij}$ are $\cO(1)$ coefficients.
With this definition, the $\Delta F=2$ Hamiltonian
can be written as
\be
\cH^{\Delta F=2}_{\rm eff} = \frac{1}{2} \left( \hat X_{LL}^2 + \hat X_{LL}^{\dagger2} \right)
= \left. \hat X_{LL}^2 \right|_{a_i\to \Re(a_i),~ b_{31}=b_{13}^*}
\label{eq:hermiti}
\ee
where $X_{ij}$ assumes the following form in
the mass-eigenstate basis of down-type quarks: 
\be
X^d = U_{d_L}^* \times 
\left( \begin{array}{ccc}  a_1 & 0 & 0 \\ 0 & a_1 + a_2 \epsilon^2  & b_{31} \epsilon \\ 
0 & b_{13} \epsilon  & a_3 \end{array} \right) \times~U_{d_L}^T~.
\label{eq:XLL1}
\ee

The $X^d$ entry relevant to kaon physics is
\bea
X^{d}_{12} & = & s_d c_d e^{-i(\phi+\delta)} \left[ s_b^2 (a_3-a_1)  -
b_{31} s_b \epsilon e^{i\phi_b}  - b_{13} s_b \epsilon
 e^{-i\phi_b } + a_2 \epsilon^2 \right] + \cO(s_d\epsilon^3)  \\
 &=& c_K V_{td}^* V_{ts} + \cO(s_d\epsilon^3)~.
\eea
Once we take into account the conditions on $a_{i}$ and 
$b_{ij}$ dictated by the hermiticity of the 
$\Delta F=2$ Hamiltonian in (\ref{eq:hermiti}), it follows that 
the $\cO(1)$ coefficient $c_K$ is real.
Similarly, the entries relevant to $B$ physics are
\bea
X^{d}_{13} &=& s_d e^{-i(\phi+\delta)} \left[
 - s_b (a_{3}-a_{1}) e^{-i ( \xi+\phi_b )}
 + b_{31}  e^{-i\xi}  \right]  + \cO(s_d \epsilon^2) 
 = c_B e^{i\alpha_B} V_{td}^* V_{tb}  + \cO(s_d \epsilon^2)~, \nonumber
 \\
X^{d}_{23} &=&
 c_d \left[  - s_b (a_{3}-a_{1}) e^{-i ( \xi+\phi_b )}
 + b_{31}  e^{-i\xi}  \right]  + \cO(\epsilon^2)  = c_B e^{i\alpha_B} V_{ts}^* V_{tb}  
+ \cO(\epsilon^2)~, 
\eea
From these structures we can generalize the following three statements
on the corrections to $\Delta F=2$ amplitudes
that we already found in the specific case of the gluino-mediated
amplitudes:
\begin{itemize}
\item[i.] in all cases the size of the correction is proportional to the 
CKM combination of the corresponding SM amplitude (MFV structure);
\item[ii.] the proportionality coefficient is the same in $B_d$ and 
$B_s$ systems, while it maybe be different in the kaon system;
\item[iii.] new CP-violating phases can only appear in the $B_d$ and 
$B_s$ systems (in a universal way).
\end{itemize}  
These statements are a general consequence of the 
$U(2)^3$ flavour symmetry and its breaking pattern. 
The properties ii.~and iii.~have indeed already been discussed in the literature 
in the context of MFV in the large $\tan\beta$ limit:
the statement ii. was already discussed in~\cite{D'Ambrosio:2002ex}, 
in absence of flavour-blind phases, while the condition 
iii.~has been pointed out in~\cite{Kagan:2009bn}.
This is not surprising,
since the $U(3)^3$ group of MFV is broken to $SU(2)^3\times U(1)$ in the 
large $\tan\beta$ limit~\cite{Kagan:2009bn}. 

Since the $U(2)^3$ group is our starting point,
these conditions are naturally realized in our framework 
independently of the value of $\tan\beta$
and without the need of considering operators 
with high powers of the spurion fields.
The only model-dependent result following from the assumption of
gluino-mediated amplitudes is the sign of the contribution to $\epsilon_K$
that, as we have seen, goes in the direction required by data.
Note also that, assuming $U(2)^3$ from the beginning, we realize 
the decoupling of $B$ and $K$ physics without the need of operators
with several powers of the Yukawa couplings (contrary to Ref.~\cite{D'Ambrosio:2002ex,Kagan:2009bn}),
operators which are naturally suppressed in a weakly coupled theory such as the MSSM.

\section{Conclusions and outlook}

Motivated in part by a few difficulties that seem to appear in the current description of the flavour and CP-violation data, especially in the sector of $\Delta F=2$ observables, and in part by the absence of large deviations from the Standard Model elsewhere, we have considered in this work the possibility that the problem of $\Delta F=2$ observables be due to the emergence of long waited signals of supersymmetry in the flavour and CP-violating sectors.
To do this, we have found particularly useful to reconsider the proposal that a weakly broken $U(2)$ symmetry be operative in determining the full flavour structure of the supersymmetric extension of the SM. Among the appealing features of $U(2)$  and an advantage over the standard MFV proposal is that it allows the first two generations of sfermions to be substantially heavier than the third one, which helps to address specifically also the supersymmetric  CP problem.
A single $U(2)$ has a problem, however: 
the dominance over every other effect of the contribution to $\epsilon_K$ due to a LR operator with its chirally enhanced $K_0-\bar{K}_0$ matrix element. The solution of this problem resides in enlarging $U(2)$ to the full $U(2)_{Q}\times U(2)_{u}\times U(2)_{d}$ symmetry of the first two generations and demanding that the communication with the third generation be due to doublets under $U(2)_Q$ only. If only one such doublet is present, characteristic correlations exist between the various $\Delta F=2$ amplitudes that we have exploited to improve the consistency of the fit of the flavour and CP-violation current data. A striking confirmation of this picture can be provided by the measurement, currently under way by the LHCb collaboration with sufficient precision, of the CP asymmetry in  
 $B_s \to \psi \phi$, predicted to be positive and above its Standard Model value: $0.05 \lsim S_{\psi\phi} \lsim 0.2$.
 Furthermore, to attribute it to supersymmetry requires finding a gluino and a left-handed sbottom with masses below about $1\div 1.5$ TeV.
 
 The present study can be extended in several directions, that we briefly mention. First, although the effects in $\Delta F=2$ amplitudes are the ones of most obvious phenomenological significance at present, some effects in $\Delta F=1$ transitions will also be present, relevant to future measurements. Specifically we refer to CP violation, both due to the phases in (\ref{CKM}) and (\ref{WL}), and to possible flavour blind phases, not strongly constrained by the EDMs 
 due to the heaviness of the first sfermion generation~\cite{Barbieri:2011vn}. 
Contrary to the $\Delta F=2$ case we expect 
the signatures in the   $\Delta F=1$ sector to be more dependent on the 
details of the model, this is why we defer their analysis to a separate paper. 
Secondly, in this work no assumption is made about possible intermediate breaking patterns of $U(2)^3$, similarly to what was done in Ref.~\cite{Barbieri:1998qs} for the $U(2)$ case. This allowed to correlate  $s_u, s_d$ in (\ref{CKM}) and (\ref{WL}) to the ratios of light quark masses $m_u/m_c$ and $m_d/m_s$. A reconsideration of these attempts, appropriately corrected, in view of the current data might be useful. Finally, the problem is pending of describing a dynamical model that realizes the phenomenological $U(2)^3$ picture. Section 4 provides an example, which may be useful to study in more detail and/or  be suitably modified.

\section*{Acknowledgments}
We thank Paolo Campli and Filippo Sala for useful discussions.
This work was supported by the EU ITN ``Unification in the LHC Era'', 
contract PITN-GA-2009-237920 (UNILHC) and by MIUR under contract 2006022501.
G.I. acknowledges the support of the Technische Universit\"at M\"unchen 
-- Institute for Advanced Study, funded by the German Excellence Initiative.

\appendix

\section{Yukawa and CKM matrix in $U(2)^3$}

The transformation properties of the quark superfields under
the $U(2)_{Q}\times U(2)_{u}\times U(2)_{d}$ group are:
\begin{eqnarray}
  Q  \equiv ( Q_{1}  , Q_{2} )^{\phantom{T}}  &\sim& (\bar 2,1,1)~, \\
  u^{c}\equiv ( u_{1}^c , u_{2}^c )^T    &\sim& (1,2,1)~, \\
  d^{c}\equiv ( d_{1}^c , d_{2}^c )^T    &\sim& (1,1,2)~, 
\end{eqnarray}
while $q_{3}$, $t^c$, and $b^c$ (the third generation fields) 
are singlets. We also assume a $U(1)_b$ symmetry under which 
only $b^c$ is charged.
With such assignments, the only term allowed in the Superpotential 
in the limit of unbroken symmetry is 
\begin{equation}
 W=y_t ~ q_{3} t^c ~H_u~,
\end{equation}
where $y_t$ is the $\ord{1}$ top Yukawa coupling.

The first step in the construction of the full Yukawas lies on the introduction 
of the $U(2)^3$-breaking spurion $V$, transforming as a (2,1,1). This 
allow us to write the following up-type Yukawa matrix\footnote{~We define 
the $3\times 3$ Yukawa matrix starting from the superpotential 
$W =  q_i (Y_u)_{ij} u_j^c ~H_u~$. 
This imply the following SM (non-supersymmetric) 
Yukawa interaction $\cL =  \bar q_{Li} (Y_u^*)_{ij} u_{Rj} ~H_c~$}
\begin{equation}
Y_u =y_t \left(\begin{array}{c:c}
 0 & x_t\,V \\\hdashline
 0 & 1
\end{array}\right).
\end{equation}
Here and in the following everything above the horizontal dashed 
line is subject to the $U(2)_Q$ symmetry, 
while everything to the left of the vertical dashed line is subject 
to the $U(2)_{u}$ symmetry (or the $U(2)_{d}$ symmetry in the down-type
sector). The parameter $x_t$ is a complex  free parameter of $\ord{1}$.

Similarly we can write the 
following down-type Yukawa matrix
\begin{equation}
Y_d= y_b \left(\begin{array}{c:c}
 0 & x_b\,V \\\hdashline
 0 & 1
\end{array}\right),
\end{equation}
where again  $x_b$ is a complex free parameter of 
$\ord{1}$. The size of $y_b$
depends on the ratio of the two Higgs VEV's. If $\tan\beta = \langle H_u \rangle/\langle H_d \rangle = \cO(1)$ the smallness of $y_b$ can be attributed to approximate $U(1)$'s inside and outside  $U(2)^3$. Otherwise we can consider as reference value  $\tan\beta = \langle H_u \rangle/\langle H_d \rangle = \cO(10)$, such that $y_b$ is small enough to avoid dangerous large $\tan\beta$ effects,
but is much larger than the $U(2)^3$ breaking spurions
and can be used as a natural overall normalization factor for the 
down-type Yukawa coupling. 


Finally, in order to build the masses and mixing of the first two generations we introduce two 
additional spurions, $\Delta Y_u$ and $\Delta Y_d$, transforming as $(2,\bar 2,1)$
and $(2,1,\bar 2)$, respectively. Combining the various symmetry breaking terms, 
the Yukawa matrices end up with the following pattern:
\begin{align}
Y_u= y_t \left(\begin{array}{c:c}
 \Delta Y_u & x_t\,V \\\hdashline
 0 & 1
\end{array}\right), & &
Y_d= y_b \left(\begin{array}{c:c}
 \Delta Y_d & x_b\,V \\\hdashline
 0 & 1
\end{array}\right),
\end{align}
where we have absorbed $\ord{1}$ couplings 
by redefining $\Delta Y_u$ and $\Delta Y_d$.

Due to the holomorphicity of the Superpotential, in a supersymmetric framework 
we are not be able to add term on the lower-left sector of the Yukawas. 
Such terms would indeed have a structure of the type
\be
 q_{3} \left(V^\dagger\,\Delta Y_u\right)\,U^{c}, \qquad 
 q_{3} \left(V^\dagger\,\Delta Y_d\right)\,D^{c}.
\ee
Beside being non-holomorphic, these terms are doubly suppressed. 
Although we will not include them in the following, 
we have explicitly checked that their 
inclusion do not lead to significant differences in the results presented 
below.

\subsection{Explicit Parametrization}
\label{sect:exppar}

The leading spurion $V$ can always be decomposed as
\begin{equation}
 V=\epsilon\,U_V  \hat s_2~, \qquad 
s_2 = \left(\begin{array}{c} 0 \\ 1 \end{array}\right)~,
\end{equation}
where $U_V$ is a $2\times 2$ unitary matrix [det($U$)=1]
and $\epsilon$ is a real parameter that we require to be of $\cO(|V_{cb}| \approx 4\times 10^{-2})$.
The $\Delta Y_u$ and $\Delta Y_d$ spurions can be decomposed as:
\begin{eqnarray}
 \Delta Y_u &=&  U_{Q_u}^\dagger \Delta Y_u^d\,U_{U}, \\
 \Delta Y_d &=&  U_{Q_d}^\dagger \Delta Y_d^d\,U_{D},
\end{eqnarray}
where $\Delta Y_u^d=\textrm{diag}(\lambda_{u1}, \lambda_{u2})$, 
$\Delta Y_d^d=\textrm{diag}(\lambda_{d1}, \lambda_{d2})$, 
and the $U$'s are again $2\times2$ unitary matrices.
By construction, the size of the $\lambda_i$ is such that 
the largest entry is $|\lambda_{d2}| \approx m_s/m_b =\cO(\epsilon)$.

With a suitable rotation in the $U(2)^3$ space we can get rid of $U_V$, $U_{U}$, and $U_{D}$.
In such base the Yukawa matrices assume the explicit form
\begin{eqnarray}
\label{Yudef}
Y_u&=& y_t \left(\begin{array}{c:c}
 U_{Q_u}^\dagger \Delta Y_u^d & \epsilon\, x_t \hat s_2 \\\hdashline
 0 & 1
\end{array}\right), \\
\label{Yddef}
Y_d&=& y_b \left(\begin{array}{c:c}
 U_{Q_d}^\dagger\Delta Y_d^d & \epsilon\, x_b \hat s_2 \\\hdashline
 0 & 1
\end{array}\right)~.
\end{eqnarray}

We shall now address the issue of the relevant CP phases. 
We first note that shifting the phases of $t^c$ and $b^c$ 
we can get rid of the phases in $y_t$ and  $y_b$, 
while a rephasing of the components of $u^{c}$ and $d^{c}$ allows
us to set the diagonal entries in $\Delta Y_{u,d}^d$ to be real. 
In principle, we can get rid of one of the two phases in 
$x_t$ or $x_b$. However, in order to maintain a symmetric notation
for up- and down-quark Yukawas, we keep them both 
complex and denote them by $x_{f} e^{i\phi_{f}}$, 
with $x_{f}$ being real and positive ($f=t,b$).
Without further rephasing we are also left with the two phases 
in $U_{Q_{u,d}}$, that we parametrize as
\begin{equation}
 U_{Q_f}=
\left(\begin{array}{cc}
c_f & s_f\,e^{i\alpha_f} \\
-s_f\,e^{-i\alpha_f} & c_f
\end{array}\right).
\end{equation}
In the following we assume that $s_f \ll 1$,
as naturally implied by some alignment of the $\Delta Y_{u,d}$ spurions 
in the $U(2)_{Q}$ space with respect to the leading $(2,1,1)$ 
breaking term.

\subsection{Diagonalization and CKM}

The Yukawas are diagonalized by  
\bea
 U_{uL} Y_u U_{uR}^\dagger &=& {\rm diag}(y_u,y_c,y_t)~ \no\\
 U_{dL} Y_d U_{dR}^\dagger &=& {\rm diag}(y_d,y_s,y_b)~.
\eea
To a good approximation, left-handed up-type diagonalization matrix is 
\bea
&&  U_{uL} =  \left(\begin{array}{c:c}
 U_{Q_u} & 0 \\\hdashline 0 & 1 
 \end{array}\right) \times  R_{23} (s_t; \phi_{t} )
\no \\
&& =  \left(\begin{array}{ccc}
 c_u &
 s_u\,e^{i\alpha_u}  & -s_u s_t e^{i (\alpha_u +\phi_t)}  \\
-s_u\,e^{-i\alpha_u} &  c_u c_t & -c_u s_t  e^{i\phi_{t}}   \\
 0   & s_t  e^{-i\phi_{t}} & c_t
\end{array}\right), 
\eea
where $s_t/c_t = \epsilon\, x_t$, and similarly for the
down-type sector (with $s_u,c_u \to s_d,c_d$, $x_t e^{i\phi_t} \to x_b e^{i\phi_b}$).
These expressions are valid up to relative corrections of 
order $\lambda_{u2} (\lambda_{d2})$ to the 1-2 and 2-3 elements of 
$U_{uL}(U_{dL})$, and even smaller corrections to the 1-3 elements. 

Contrary to the left-handed case, the right-handed diagonalization 
matrices become the identity in the limit of vanishing light-quark masses
(or vanishing $\Delta Y_{u,d}^d$). Neglecting the first generation
eigenvalues, and working to first order in $\epsilon\lambda_{u2}$
and  $\epsilon\lambda_{d2}$, we get
\bea
U_{uR} &=& \left(\begin{array}{ccc}
 1 & 0 & 0  \\
 0 & 1 & -\lambda_{u2} s_t  e^{i\phi_{t}} \\
 0 &  \lambda_{u2} s_t e^{-i\phi_{t}} & 1
\end{array}\right)~, \no \\
U_{dR} &=& \left(\begin{array}{ccc}
 1 & 0 & 0  \\
 0 & 1 & -\lambda_{d2} s_be^{i\phi_b} \\
 0 &  \lambda_{d2} s_be^{-i\phi_b} & 1
\end{array}\right)~. 
\eea

We are now ready to evaluate the CKM matrix $V_{\rm CKM}= (U_{uL}\cdot U_{dL}^\dagger)^*$. 
Using the decomposition above we find 
\bea
V^{(0)}_{\rm CKM} &=&
 \left(\begin{array}{c:c}
 U^*_{Q_u} & 0 \\\hdashline 0 & 1 
 \end{array}\right) \times  R_{23} (s;\xi) 
\times 
 \left(\begin{array}{c:c}
 U^T_{Q_d} & 0 \\\hdashline 0 & 1 
 \end{array}\right)~, \\
&\approx& \left(\begin{array}{ccc}
 c_u c_d + s_us_d\,e^{i(\alpha_d-\alpha_u)}  &
 -c_u s_d \,e^{-i\alpha_d}  +s_u c_d \,e^{-i\alpha_u}    & s_u s e^{- i (\alpha_u-\xi) } \\
 c_u s_d \,e^{i\alpha_d}  - s_u c_d \,e^{i\alpha_u}    &  c_u c_d + s_us_d\,e^{i(\alpha_u-\alpha_d)}  &
 c_u s  e^{i\xi} \\
-s_d s \,e^{i (\alpha_d-\xi)} & -s c_d  e^{-i\xi} & 1 \\
\end{array}\right)~. \label{eq:long1} 
\eea
where $(s/c) e^{i\xi} = \epsilon x_b e^{-i\phi_{b}} -\epsilon x_t e^{-i\phi_{t}} $, 
and where we have set $c=1$.
With an  appropriate rephasing of the fields this structure is equivalent to the one 
in Ref.~\cite{Barbieri:1998qs}
(with $\phi=\alpha_d-\alpha_u$ and $s_{u,d} \to -s_{u,d}$). To match this structure 
with the standard CKM parametrization, we rephase it imposing real 
 $V_{ud}$, $V_{us}$, $V_{cb}$, $V_{tb}$, and  $V_{cs}$ (which is real at the 
level of approximation we are working), obtaining 
\be
 V_{\rm CKM}=\left(\begin{array}{ccc}
 1- \lambda^2/2 &  \lambda & s_u s e^{-i \delta}  \\
-\lambda & 1- \lambda^2/2   & c_u s  \\
-s_d s \,e^{i (\phi+\delta)} & -s c_d & 1 \\
\end{array}\right),
\label{eq:CKMstand}
\ee
where $\phi = \alpha_d - \alpha_u$, while the phase $\delta$ and 
the  real and positive parameter $\lambda$ are defined by 
\be
s_uc_d - c_u s_d e^{-i\phi}  = \lambda e^{i \delta}.
\ee

The parametrization (\ref{eq:CKMstand})  
is equivalent, in terms of precision, to the 
Wolfenstein parametrization up to $\cO(\lambda^4)$ and, similarly to the latter, 
can be systematically improved considering higher powers in $s$, $s_d$, and $s_u$. 
The four parameters $s_u$, $s_d$, $s$, and $\phi$ can be determined 
completely (up to discrete ambiguities) in terms of the four independent 
measurements of CKM elements. In particular, using tree-level 
inputs we get  
\bea
s  &=& |V_{cb}| = 0.0411 \pm 0.0005~, \\
\frac{s_u}{c_u} &=& \frac{|V_{ub}|}{|V_{cb}|} =  0.095  \pm 0.008~, \\
s_d &=&  - 0.22  \pm 0.01 \qquad {\rm or} \qquad -0.27 \pm 0.01~.
\eea
As a consequence of the $U(2)_Q$ symmetry, 
$|V_{td}/V_{ts}|$ is naturally of $\cO(\lambda)$ and 
the smallness of $|V_{ub}/V_{td}|$
is attributed to the smallness of $s_u/s_d$~\cite{Barbieri:1997tu}.
The latter hypothesis fits well, at least qualitativey, with 
the strong alignement of the spurions $\Delta Y_u$ and $V$ in the  
$U(2)_Q$  space indicated by the smallness of $m_u/m_c$.

\section{Soft-breaking masses}

In the limit of unbroken symmetry the three soft mass matrices have  the following structure:
\begin{equation}
m^2_{\tilde f}=\left(\begin{array}{ccc}
m_{f_h}^2 & 0 & 0 \\ 
0 & m_{f_h}^2 & 0 \\ 
0 & 0 & m_{f_l}^2 \end{array}\right) 
\end{equation}
where the $m_{f_i}^2$ and are real parameters. 
When the spurions are introduced in order to build the Yukawas, they also affect 
the structure of the soft masses, which assume the form
\bea
m^2_{\tilde Q} &=&  m_{Q_h}^2
\left(\begin{array}{c:c} 
1  + \Delta_{LL}
&   x_{Q} e^{-i\phi_Q} V^* \\ \hdashline
    x_{Q} e^{i\phi_Q} V^T  \phantom{A^{A^{A^A}}}  &   m_{Q_l}^2/m_{Q_h}^2 
\end{array}\right)~,  \\
m^2_{\tilde d} &=&  m_{d_h}^2
\left(\begin{array}{c:c}
 1  +  c_{dd}  \Delta Y_d^T \Delta Y_d^{*}
&  x_{d} e^{-i\phi_d} \Delta Y_d^T V^* \\ \hdashline
   x_{d} e^{i\phi_d} V^T \Delta Y_d^{*}  \phantom{A^{A^{A^A}}}  &   m_{d_l}^2/m_{d_h}^2 
\end{array}\right)~,  \no \\
m^2_{\tilde u} &=& m_{u_h}^2
\left(\begin{array}{c:c} 
 1  +  c_{uu}  \Delta Y_u^T \Delta Y_u^{*}
&    x_{u} e^{-i\phi_u} \Delta Y_u^T V^* \\ \hdashline
     x_{u} e^{i\phi_u} V^T \Delta Y_u^{*}  \phantom{A^{A^{A^A}}}  &   m_{u_l}^2/m_{u_h}^2 
\end{array}\right)~, \no 
\eea
where 
$$
\Delta_{LL} = 
c_{Qv} V^* V^T   +  c_{Qu}  \Delta Y_u^* \Delta Y_u^{T} 
+  c_{Qd}  \Delta Y_d^* \Delta Y_d^{T} 
$$
and $c_i,x_i$ are real $\cO(1)$ parameters.

Let's consider first the case of $m^2_{\tilde Q}$. 
In the limit where we neglect light quark masses ($\Delta Y_{u,d} \to 0$), 
adopting the explicit parametrization in sect.~\ref{sect:exppar}, 
we have
\bea
&&  R_{23}(s_Q; -\phi_Q) \times m^2_{\tilde Q} 
\times R_{23}(- s_Q; -\phi_Q) \no \\
&&  = (m^2_{\tilde Q})^d = {\rm diag}( m_{Q_1}^2, m_{Q_2}^2, m_{Q_3}^2 )
\eea
where  $s_Q/c_Q  = \epsilon x_{Q}/(1- m_{Q_l}^2/m_{Q_h}^2) \approx  \epsilon x_{Q}$ and 
\bea
m_{Q_1}^2 &=&  m_{Q_h}^2~, \\
m_{Q_2}^2 &=&  m_{Q_h}^2 \left(1+ \epsilon^2 c_{Qv}+  \epsilon^2 x^2_{Q}\right) 
+\cO( \epsilon^2 m_{Q_l}^2)~, \\
m_{Q_3}^2 &=&  m_{Q_l}^2 -\epsilon^2 x_{Q} m_{Q_h}^2  +\cO( \epsilon^2 m_{Q_l}^2 )~.
\eea
This implies that in the mass-eigen\-state basis of down quarks,  $m^2_{\tilde Q}$ 
is  diagonalized by 
\bea
W_L^{d\dagger} ~m^2_{\tilde Q}~ W^d_L &=&  {\rm diag}( m_{Q_1}^2, m_{Q_2}^2, m_{Q_3}^2 )~, \\
W^d_L &=& U_{d_L}^* \times  R_{23}(- s_Q; -\phi_Q)~.
\eea
With such definition the coupling of the gluinos to left-handed 
down-type quarks and squarks in their mass-eigen\-state basis is 
governed by $[\bar d^i_L (W^d_L)_{ij}\, \tilde q^j_L]\, \tilde g$. 

Employing the CKM phase convention in (\ref{eq:CKMstand}) 
for both down-type quarks and
down-type squarks, the mixing matrix $W^d_L$
assumes the  form
\bea
W^d_L &=& \left(\begin{array}{ccc}
 c_d &  s_d  e^{-i(\delta +\phi)}  & -s_d s_L e^{i\gamma} e^{-i(\delta +\phi)}  \\
-s_d e^{i(\delta +\phi)}  &  c_d & -c_d s_L e^{i\gamma}   \\
  0  &  s_L e^{-i\gamma} & 1 \\
\end{array}\right) \no\\
&=& \left(\begin{array}{ccc}
 c_d &  \kappa^* & - \kappa^*  s_L e^{i\gamma}  \\
- \kappa  &  c_d & -c_d s_L e^{i\gamma}   \\
  0  &  s_L e^{-i\gamma} & 1 \\
\end{array}\right),
\eea
where  $\kappa =  c_d V_{td}/V_{ts}$,
\bea
s_L e^{i\gamma}  &=&   e^{-i\xi} (s_{x_b} e^{ - i \phi_b}  + s_Q e^{ - i \phi_Q}) \no\\
&\approx&
\epsilon  x_b e^{-i(\xi+\phi_b)} \left( 1+ \frac{x_{Q}}{x_b} e^{i(\phi_b - \phi_Q)} \right)~,
\label{eq:gamma}
\eea 
and, as usual, we have neglected $\cO(\epsilon^2)$ corrections.  
To understand the structure of $i\to j$ FCNCs, it is useful to 
consider the combinations 
\be
\lambda^{(a)}_{i\not = j} = (W^d_L)_{ia} (W^d_L)^*_{ja}~, \qquad 
\lambda^{(1)}_{ij} + 
\lambda^{(2)}_{ij} + 
\lambda^{(3)}_{ij} =0~,
\ee
for which we find
\be
\begin{array}{ccclclcl} 
\lambda^{(2)}_{ij} & = & 
c_d \kappa^* +\cO(s_L^2 \kappa^*) & [ {}_{ij=12}], &  +s_L \kappa^* e^{i\gamma}  & [ {}_{ij=13}],
&  +c_d s_L e^{i\gamma}  & [ {}_{ij=23}],   \\ 
\lambda^{(3)}_{ij} & = & 
s_L^2 \kappa^* c_d  & [ {}_{ij=12}],  & -s_L \kappa^* e^{i\gamma} & [ {}_{ij=13}],
&  -c_d s_L e^{i\gamma}  & [ {}_{ij=23}]. \\
\end{array}
\ee
From these structures we deduce that $2\to 1$ transitions receive contributions 
aligned, in phase, with respect to the SM term. 
The new non-trivial phase $\gamma$ enters only in 
FCNC transitions of the type $3\to 1,2$, where it appears 
as a universal correction relative to the CKM phase. 
Note that the phase $\gamma$ and the $2\to 3$ effective 
mixing angle $s_L$ do not vanish also in the limit of vanishing 
breaking terms in the soft mass matrix ($x_Q\to 0$). 

Switching-on the $\Delta Y_{u,d}$ terms in $m^2_{\tilde Q}$
leads to tiny corrections to  $W_L^d$ that
can be neglected to a good approximation, similarly to the 
light-quark corrections to the CKM matrix elements. 
The most significant impact of  $\Delta Y_{u,d}\not=0$
is on the mass splitting of the first two generations. 
In principle, the mass splitting of the first two generations
has a non-negligible impact on $2\to 1$ FCNC transitions (Kaon physics). 
However, in the limit $m_{Q_l}^2 \ll m_{Q_h}^2$ 
it becomes negligible also for  $2\to 1$ transitions.

\bibliographystyle{My}
\bibliography{u2mfv}

\end{document}